\documentclass[twoside,twocolumn]{article}

\usepackage{blindtext} 
\usepackage{textcase}

\usepackage[sc]{mathpazo} 
\usepackage[T1]{fontenc} 
\usepackage{microtype} 

\usepackage[english]{babel} 

\usepackage[hmarginratio=1:1,top=32mm,columnsep=20pt]{geometry} 
\usepackage[hang, small,labelfont=bf,up,textfont=it,up]{caption} 
\usepackage{booktabs} 

\usepackage{lettrine} 

\usepackage{enumitem} 
\setlist[itemize]{noitemsep} 

\usepackage{abstract} 

\usepackage{titlesec} 
\renewcommand\thesection{\Roman{section}} 
\renewcommand\thesubsection{\roman{subsection}} 
\titleformat{\section}[block]{\large\scshape\centering}{\thesection.}{1em}{} 
\titleformat{\subsection}[block]{\large}{\thesubsection.}{1em}{} 

\usepackage{fancyhdr} 
\usepackage{titling} 

\usepackage{hyperref} 

\usepackage{tabu}                      
\usepackage{booktabs}                  
\usepackage{multirow}

\usepackage{enumitem}
\usepackage{amsfonts}
\usepackage{wrapfig}
\usepackage{float}
\usepackage{subfig}

\usepackage{graphicx}
\usepackage{citeall}
\usepackage{textcomp}

\pretitle{\begin{center}\Huge\bfseries} 
\posttitle{\end{center}} 
\title{Navigating a maze differently - a user study} 
\author{%
\textsc{Aryabrata Basu}\thanks{Corresponding author} \\[1ex] 
\normalsize Emory University \\ 
\normalsize \href{mailto:aryabrata.basu@emory.edu}{aryabrata.basu@emory.edu} 
\and 
\textsc{Kyle Johnsen}\thanks{Advisor} \\[1ex] 
\normalsize University of Georgia \\ 
\normalsize \href{mailto:kjohnsen@uga.edu}{kjohnsen@uga.edu} 
}
\date{\today} 
\renewcommand{\maketitlehookd}{%
\begin{abstract}
\noindent Navigating spaces is an embodied experience. Examples can vary from rescue workers trying to save people from natural disasters; a tourist finding their way to the nearest coffee shop, or a gamer solving a maze. Virtual reality allows these experiences to be simulated in a controlled virtual environment. However, virtual reality users remain anchored in the real world and the conventions by which the virtual environment is deployed influence user performance. There is currently a need to evaluate the degree of influence imposed by extrinsic factors and virtual reality hardware on its users. Traditionally, virtual reality experiences have been deployed using Head-Mounted Displays with powerful computers rendering the graphical content of the virtual environment; however, user input has been facilitated using an array of human interface devices including Keyboards, Mice, Trackballs, Touchscreens, Joysticks, Gamepads, Motion detecting cameras and Webcams. Some of these HIDs have also been introduced for non-immersive video games and general computing. Due to this fact, a subset of virtual reality users has greater familiarity than others in using these HIDs. Virtual reality experiences that utilize gamepads (controllers) to navigate virtual environments may introduce a bias towards usability among virtual reality users previously exposed to video-gaming.

This article presents an evaluative user study conducted using our ubiquitous virtual reality framework with general audiences. Among our findings, we reveal a usability bias among virtual reality users who are predominantly video gamers. Beyond this, we found a statistical difference in user behavior between untethered immersive virtual reality experiences compared to untethered non-immersive virtual reality experiences. 
\end{abstract}
}

\begin{document}

\maketitle


\section{Introduction}

\lettrine[nindent=0em,lines=3]{S}
{\MakeTextLowercase{p}}atial navigation is one of the core abilities of human beings. It is a useful skill set that we employ on a daily basis to navigate building interiors or busy streets to reach our destinations. The act of navigating a physical environment requires the simultaneous operation of both cognitive and motor functions \cite{SPIERS20061826}. The majority of virtual reality (VR) experiments are designed with spatial navigation in mind as the primary means for exploring virtual environments (VEs) \cite{WERKHOVEN2014110}. That said, the potential implications of VE interface and VR task model design on individual user performance in VR have yet to be explored adequately. As a result, understanding the underlying concepts of spatial navigation as a mathematical construct and its relationship to immersive VR user performance is particularly important, especially if we were to adopt VR exercises designed to promote navigation skill set building in users with measurable gain in output.

In this work, we presume that the prior video-gaming experience, age and the gender of the users participating in a VR study impacts their respective performance in executing immersive VR tasks. This study has been extended from earlier work examining how physiological factors affect user performance in immersive VR \cite{7460057}. In this article, we describe our virtual maze application as well as the detailed system design. We report an extended analysis conducted on the recorded user trajectory data. To that end, we defined a set of mathematically derived features generalized for each trajectory such as distance traveled, decision points reached inside the maze, positional curvature, head rotation amount, and coverage of the maze.



\begin{figure}[t]
 \centering
 \includegraphics[width=\columnwidth]{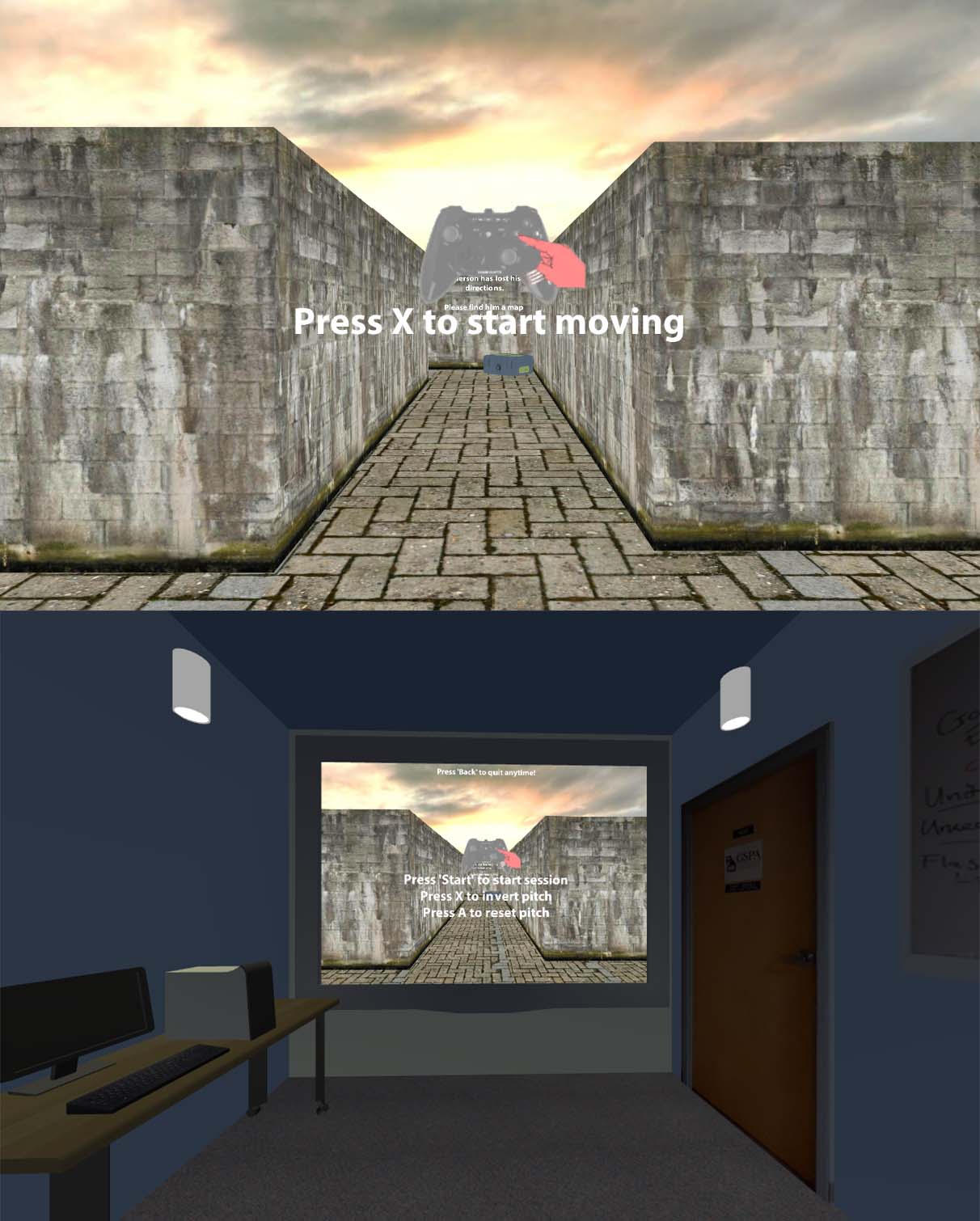}
 \caption{Immersive (top) and non-immersive (bottom) perspective of all study participants at the start of the maze experience}
 \label{fig:BothPerspective}
\end{figure}

\begin{figure}[t]
 \centering 
 \includegraphics[width=\columnwidth]{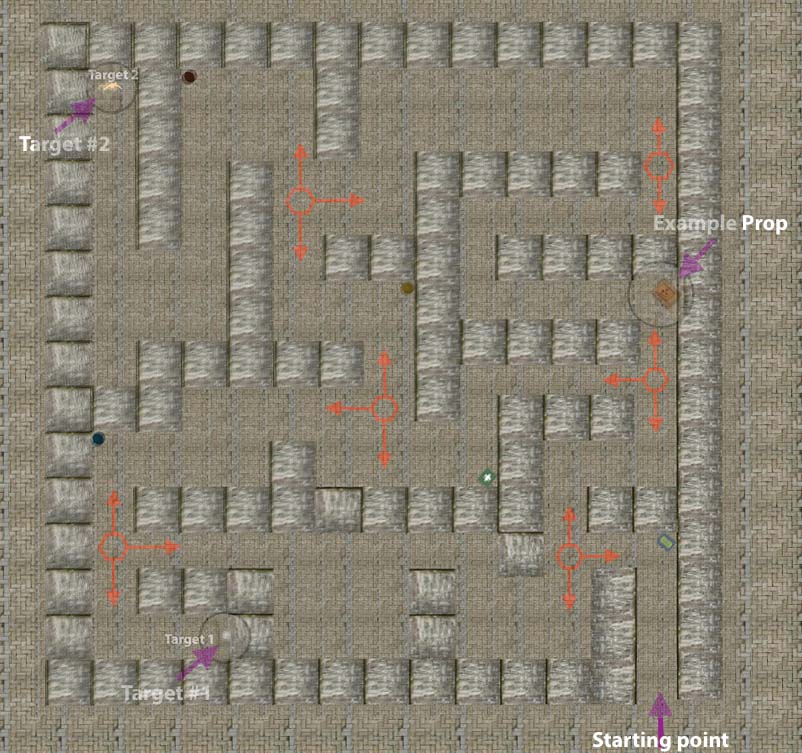}
 \caption{Key locations and decision nodes of the maze}
 \label{fig:m_anatomy}
\end{figure}

\section{Related Work}\label{sec:relatedWork}
Our application and system is built upon the strategies, technology, and research involved in previous studies conducted on the topic of spatial navigation in immersive VR. A number of researchers have addressed issues related to spatial navigation and travel metaphors in immersive VEs over the years \cite{Bowman1998,suma2010effects,suma2010evaluation,zanbaka2005comparison,jeong2005differentiation}. A subset of these works have also looked into the affects of prior video gaming experience on immersive VR navigation performance \cite{smith2009measuring}. 

In 1998, Bowman et al. proposed a formalized methodology and framework \cite{Bowman1998} for the evaluation of travel techniques in immersive VEs. The basic construct of their framework was a taxonomy of travel techniques. Their experimental analysis revealed the need to gather more information about user analytics inside VEs. In a second study, Bowman et al. studied the effects of various travel techniques on the spatial orientation \cite{Bowman1999} of users inside immersive VR environments. At the same time, Ruddle et al. introduced a formal study comparing HMDs and desktop displays \cite{ruddle1999navigating}. The objective of this study was to perform a baseline investigation that compared the two different types of displays. They found that participants using the HMD navigated the virtual buildings significantly more quickly, and developed a significantly accurate sense of relative straight-line distance. Behavioral analyses showed that participants took advantage of the natural, head-tracked interface provided by the HMD in ways that include ``looking around'' more often while traveling through the VE and spending less time being stationary while choosing a direction in which to travel. 

In 2009, Smith et al. conducted a research study to look into the impact of previous computer gaming experience, user perceived gaming ability, and actual gaming performance on navigation tasks in a VE \cite{smith2009measuring}. They found that perceived gaming skill and progress in a linear first person shooter (FPS) game were found to be the most consistent metrics. Both perceived gaming skill and progress in a linear FPS game bore a relationship to performance in trivial searches, primed searches, the number of mistakes when performing an advanced travel technique (jumping) and in traveling time requiring high speed and accuracy. Smith et al. \cite{smith2009measuring}, stated `this may require the development of a gamer profile including both similar metrics to those used in this paper, metrics over other 3D interface tasks such as selection and manipulation and metrics for subjective conditions such as user disorientation, cybersickness and presence.' In 2010, Suma et al. reported a user study comparing real walking with three virtual travel techniques; namely, gaze-directed, pointing-directed, and torso-directed travel \cite{suma2010effects}. Suma et al. found that real walking is superior in terms of user performance as it allowed more cognitive capacity for processing and encoding stimuli than pointing-directed travel metaphors. They also found that male participants were slower and performed significantly worse on the attention task when the spatial task became more difficult in contrast to female participants. In another related study, Suma et al. reported that for complex VEs with numerous turns, virtual travel techniques are acceptable substitutes for real walking if the goal of the application involves learning or reasoning based on information presented in the environment \cite{suma2010evaluation}. 

Another component of studying spatial navigation is spatial trajectory analysis. In 2005, Zanbaka et al. described a between-subjects experiment that compared four different methods of travel, their effect on cognition, and paths taken in an immersive VE \cite{zanbaka2005comparison}. This study used participants' trajectory (position and head orientation) data in post-analysis by creating overlays that ultimately revealed further differences in travel techniques. This study favored a large tracked space over other travel techniques in VR for applications where problem solving and interpretation of material is important or where opportunity to train is minimal. Jeong et al. in 2005, reported on the differentiation of information-gathering ability in the real and the VE \cite{jeong2005differentiation}. An important finding of their study was that the users path of finding information was similar, their information gathering ability differs between the real and the VE.

This article does not address the cognitive issues related to spatial exploration on an individual basis. Rather our approach distinguishes task performance (as a trend) between two categories of users (gamers and non-gamers) using VR systems. More than one performance metric for solving a maze in VR has been deployed, which allowed for a nuanced discussion of navigation abilities.

Prior work on the evaluation of ubiquitous, smartphone driven 3DUIs helped guide our system architecture~\cite{6184191,6550213}. A light-weight VR system is logistically a sound design when it comes to recruiting higher number of subjects for a research study. Other studies exist that looked into physical activity as a potential marker for performances in immersive VR environments. One such study suggested that physical activities of users can be a predictor for VR task performance~\cite{6184191}. In regards to physical activity, virtual reality has also been explored as a means to motivate individuals to exercise, a concept better known as VR exergaming~\cite{6777460}. While all these studies explore the relationship between physical activity and VR performance, there has been a lack of concrete connections between them. In contrast to quantifying physical fitness of users as suggested by Basu et al.\cite{7460057,6777460}, our approach shifts the attention of the user performance predictor to the gaming profile.

\section{Application}\label{sec:application}
Our maze application requires the user to navigate a maze and find multiple target sites. For consistency, every player had a fixed starting point and fixed targets to reach in the maze. The context of the maze experience was similar to a previous study \cite{7460057}, in which the subject is required to find and rescue another human (avatar). The task included finding a clue (minimap) before finding the human avatar and then tracing back the path to the entrance of the maze. On average, the participants spent \textit{4.64} minutes in the immersive user interface setting and \textit{6.18} minutes in the non-immersive user interface setting.

\subsection{Game Design}
The focal point of our VR experience was to traverse a maze environment, find multiple target sites and trace back your path. To add to the experience, we used an altruistic framework by setting up a background story of a lost traveler inside the maze. Each participant was given the tasks of searching for and rescuing the lost character in the maze. To better emphasize this point, a restriction was placed upon the participant in the form of finding a map of the maze first. Once the user was successful in finding the map location, the user was then asked to find the lost character in the maze. Upon meeting the character, which is an animated humanoid avatar, the participants were required to trace back their path to the entrance of the maze to complete the experience. The maze design also included strategically placed unique props (box, first-aid kit, etc.) to help with user localization and memory. Various key locations, including targets and props in our maze design, are illustrated in \autoref{fig:m_anatomy}.

\subsection{System Hardware}
\begin{figure}[t]
 \centering 
 \includegraphics[width=\columnwidth]{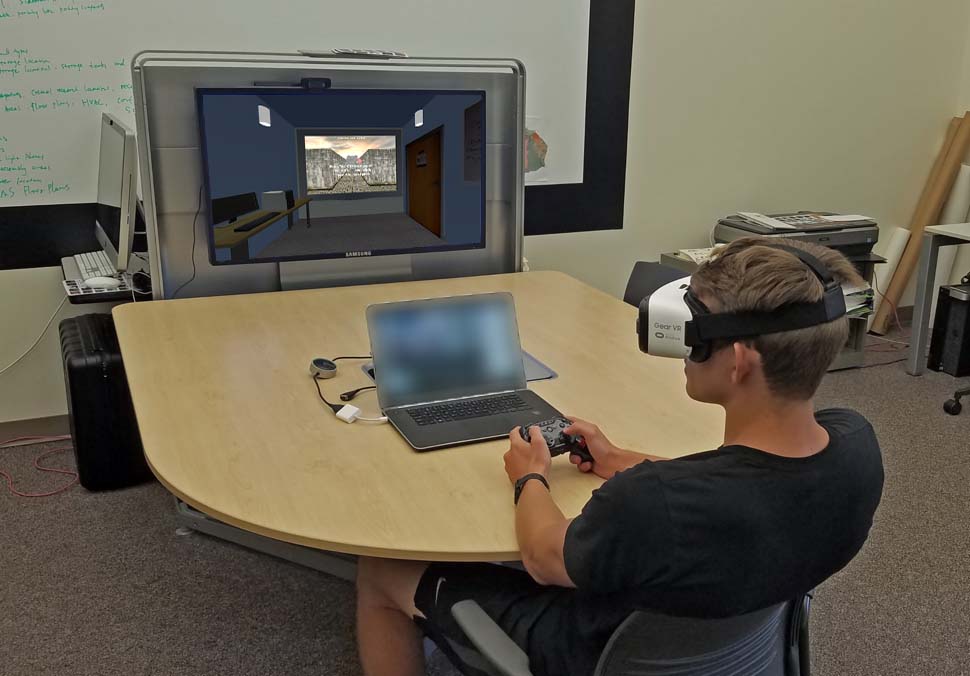}
 \caption{A randomly selected participant partaking the maze experience}
 \label{fig:participant}
\end{figure}

Our system setup included a mobile HMD and a wireless gaming controller. We opted for Samsung Galaxy Gear VR as our HMD solution and Samsung Galaxy S6 Edge+ with 5.7 inch screen size at 2560 pixels x 1440 pixels (518 ppi) screen resolution as our primary smart phone display. For user controls in the maze, we opted for a Bluetooth compatible \textit{Madcatz C.T.R.L.} gamepad controller which connects easily with our display. A typical system setup deployed for the study is illustrated in \autoref{fig:participant}.

\subsection{System Software}

\begin{figure}[t]
 \centering 
 \includegraphics[width=\columnwidth]{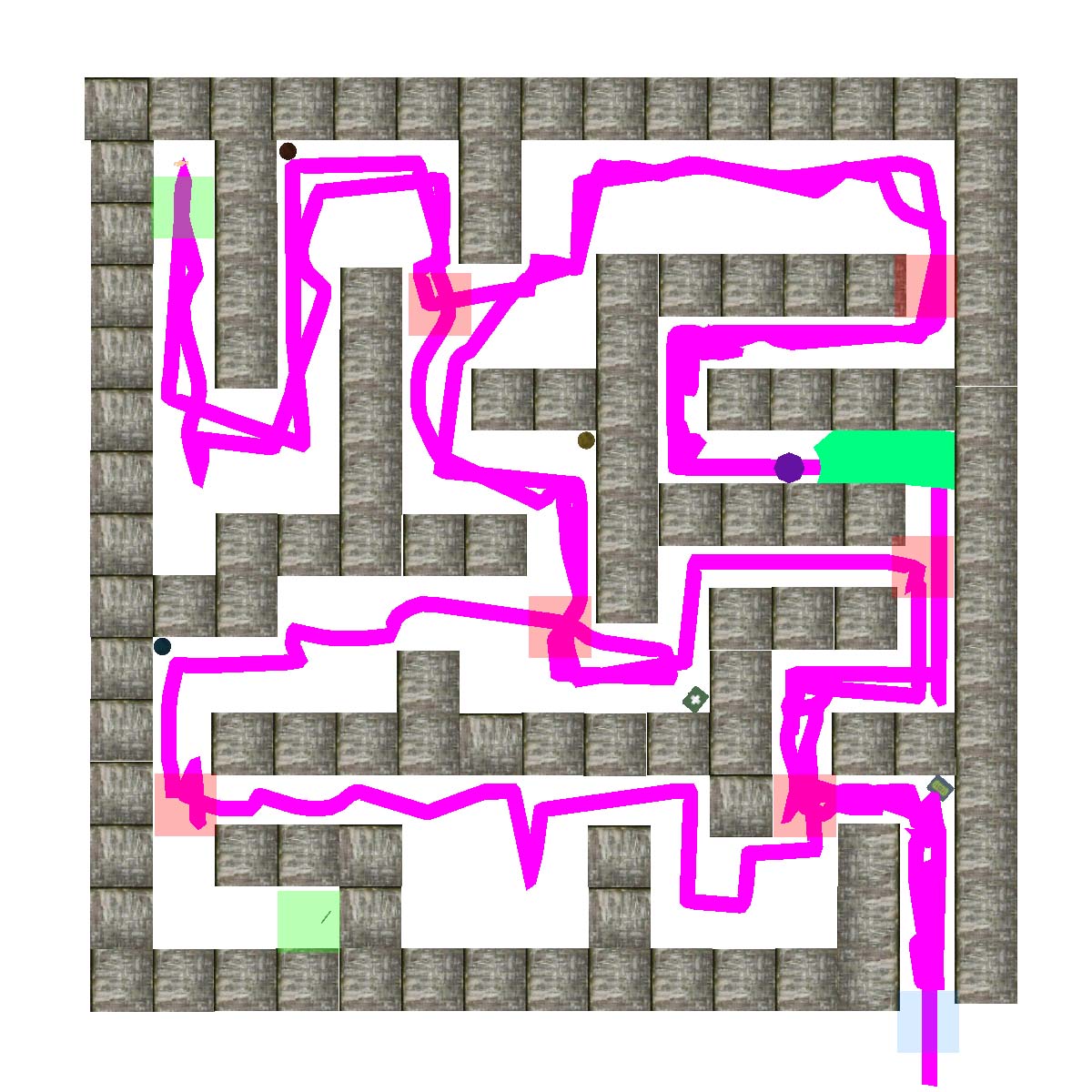}
 \caption{Visualization of a randomly selected participant's spatial trajectory data including gaze information represented as highlighted in cyan.}
 \label{fig:sampleTrajectory}
\end{figure}

The VR maze experience was developed using the Unity game engine and deployed on the HMD device as an Android app. While each participant was busy solving the maze, the system software kept an active log of their in-session trajectory data along with their head orientation data every 10 milliseconds. Furthermore, the system software also logged relevant event data such as participants quitting the maze experience or participants finding the targets successfully. A visual example of participant tracking algorithm is illustrated in \autoref{fig:sampleTrajectory}.

Once the user data was collected, scripts developed in C\# programming language were used to analyze the users' trajectory data in Unity game engine. These scripts parse through user transformation (x,y,z) in real-time and visualize their search patterns in the maze. This, in turn, allows to confirm visually emerging differences in user performance in immersive VR. This sort of visual analysis can be generalized to apply to other studies concerning spatial navigation in immersive VR.

\begin{table}
\centering
\begin{tabular}{@{}llrr@{}}\\ \toprule
& & Gamer & N-Gamer\\ \midrule
Participants & & 21 & 19\\ \midrule
\multirow{2}{*}{Gender} & Male & 14 & 4\\
& Female & 7 & 15\\ \midrule
\multirow{5}{*}{Player Level} & Novice & 0 & 12\\
& Casual & 6 & 6\\
& Pro & 12 & 0\\
& Hardcore & 2 & 0\\
& Experiential & 1 & 1\\ \bottomrule
\end{tabular}
\caption{Study demographics, based on the questionnaires; `Player Level' is a derived scale reflecting participant's involvement and approach to video-gaming; `Gamer' vs.\ `Non-Gamer'  reflects the self-assessment of being a gamer}
\label{tab:participants}
\end{table}

\section{Study Design}\label{sec:studyDesign}
\textbf{Null hypothesis}: \textit{There is not a significant impact on users' ability to successfully navigate a 3D maze under the influence of technical factors such as immersive, non-immersive user interface.}

\textbf{Alternate hypothesis}: \textit{There is a significant impact on users' ability to successfully navigate a 3D maze under the influence of technical factors such as immersive, non-immersive user interface.}

The primary goal behind the maze experience is to understand how users with varying degree of exposure to video gaming solve a complex three dimensional immersive navigation problem. We wanted to quantify spatial decision making in terms of user activity in the maze. To this end, we conducted a 2 x 1 study with user type being one of our independent variables. We recorded spatial user activity, including gaze activity, every 10 millisecond apart to help develop a model for spatial navigation performance. The experiment required users to navigate a maze with a three-fold task model. The following subsections describe the study design in details.

\subsection{Population and Environment} Over a course of 8 months, 40 self-motivated participants (mean-age = 35.93; SD = 11.11) volunteered for our study. \autoref{tab:participants} shows the breakdown of the two study groups by population demographics. These participants comprise a mix (by age, profession, background) of university staff and students at Emory University\textquotesingle s main campus. In compliance with the Institutional Review Board guidelines of both University of Georgia and Emory University, every participant\textquotesingle s consent was obtained and recorded on paper forms. Furthermore, the users were asked to fill out a series of forms including a general demographic questionnaire, pre-simulator and post-simulator sickness questionnaire and a presence questionnaire. Additionally, these users were also asked to fill out a self-reported gaming profile assessment form. Each user was then tasked to solve the same maze problem twice using an immersive user interface and then a non-immersive user interface. The order of deployment for each user was random and can be classified between the following two categories:

\begin{description}
\item[Condition 1.] Immersive first, and then non-immersive.
\item[Condition 2.] Non-immersive first, and then immersive.
\end{description}

\begin{table}
\begin{center}
    \begin{tabular}{@{}p{4.5cm}p{2.5cm}@{}}\toprule
    Features & Definition \\ \midrule
    Distance traveled & It is the total length of the path traveled between two positions. \\[0.5em] 
    Coverage & Total number of unit cubes covered (area). \\[0.6em]
    Number of decision points reached & Total number of decision points covered in the maze.  \\[0.5em]
    Positional Curvature & The signed angle of curvature between two consecutive position vectors.   \\[0.5em]
    Head rotation amount & The unsigned head rotation angle between two consecutive rotation transforms.  \\[0.5em]
    \hline
    \end{tabular}

\end{center}
\caption [User trajectory features]
{Features extracted from users' trajectory data for deep exploration and their corresponding definitions.}
\label{table:ucavecurrentmazefeaturestable}
\end{table}

Each participant was recruited using an online recruitment drive advertised through the internal university mailing list. A specific laboratory was chosen to run the sessions with only one participant at a time. The choice of our study space was particularly critical to remove any additional social anxiety within our participants. Our study investigator sat down with our participants and would start conversing to ease them into our study. Upon agreeing to our study requirements, a participant would start his/her first round of maze solving followed by a five minute break before starting his/her second round. The order of deployment for each user was random. The study investigator made sure to capture an equal number of participants for each category of deployment. Between the two sessions, the users were given five minute of break time to normalize their fatigue level before starting their next session.

\subsection{Measures} Data automatically recorded during the course of the maze navigation was used to measure the performance of each participant. For each session, we recorded, every 10 millisecond, the participant's spatial trajectory inside the maze along with orientation of gaze information (\autoref{fig:sampleTrajectory}). This allowed us to effectively playback and calculate each participant's spatio-temporal activity including, but not limited to, time spent at decision nodes in the maze and cumulative gaze rotation at decision nodes in the maze.

In addition, the participants filled out background survey questionnaires with emphasis on video gaming activity profiling, pre-simulator and post-simulator sickness questionnaire (SSQ) \cite{SSQ} and a presence questionnaire \cite{usoh2000using} related to their immersive maze experience.

\begin{figure}[th]
 \centering 
 \includegraphics[width=\columnwidth]{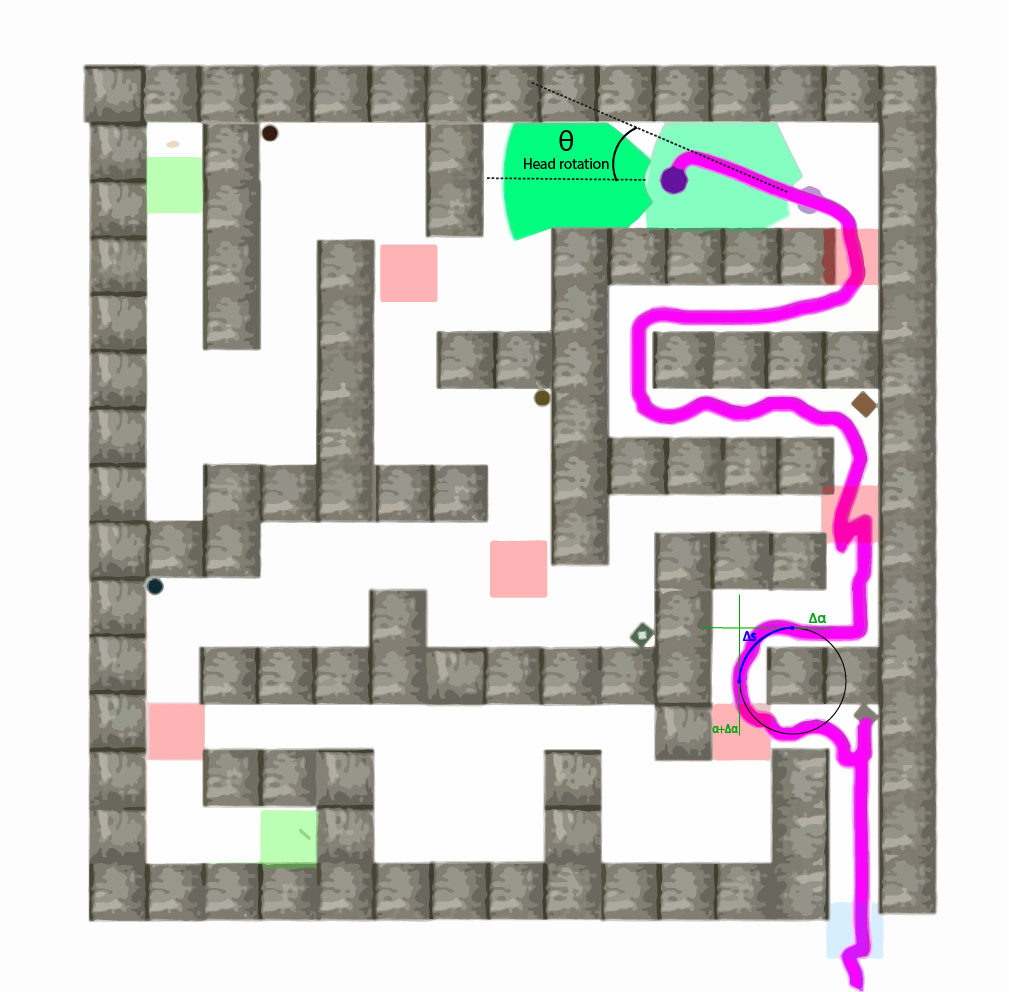}
 \caption [Visualization of two user trajectory features]
 {Visualization of two trajectory features: positional curvature and head rotation amount.}
 \label{fig:featurestrajectory}
\end{figure}

The data contained information retrieved from 40 participants with the following attributes: subject id, date, player profile, gender, age, 20/20 vision, gaming hours, athletic score, condition of deployment, time of completion, relative simulator sickness score (low: 16; high: 160), and presence score (low: -21, high: +21). Furthermore, we introduced a derived scale of attribute based on participant's self reported gaming hours per week and their video-gaming exposure profiling which we have termed as `Player Level' [see \autoref{tab:participants}]. Additionally, for deep exploration of the trajectory data, we defined a set of mathematically derived features generalized for each trajectory such as distance traveled, decision points (nodes) reached inside the maze, positional curvature, head rotation amount, and coverage of the maze. Some of these trajectory features are illustrated in Figure \ref{fig:featurestrajectory}. Positional curvature feature refers to the curvature of the trajectory calculated per frame between successive position vectors. Rotation amount feature refers to the unsigned angle calculated per frame between successive head rotation transforms stored as Quaternions. Coverage feature is simply the number of unit cubes covered (area) by the user, calculated per frame. We conducted a follow up secondary analysis of the trajectory data using the Dynamic Time Warping (DTW) algorithm \cite{bellman1959adaptive} to compute alignment distance between two trajectory samples from our pool of participant data. We used R and SPSS together to conduct the analysis of the trajectory data.

We obtained information from each sessions (immersive and non-immersive user interface). Pre-experience and post-experience survey questions relevant to this analysis are shown in \autoref{tab:questions}.

\begin{table}
\centering
\begin{tabular}{@{}p{4.8cm}p{2.5cm}@{}}\toprule
Survey Question & Response type\\ \midrule
\multirow{3}{4.8cm}{Please indicate the number of hours (per week) you play video games} & Ordinal,\newline 1 (1-3 hours) \dots\newline 5 (10+ hours)\\[.6em]
\multirow{3}{4.8cm}{Please indicate your overall (perceived) athletic skills} & Ordinal,\newline 1 (Very poor) \dots\newline 10 (Excellent)\\[.6em]

\multirow{3}{4.8cm}{Please indicate your response to the questions related to playing video games} & Ordinal,\newline Yes/No/Maybe\\[2.0em] 

\multirow{3}{4.8cm}{Please indicate and rate any of the symptoms listed in the table below on a scale of 1 to 10 (SSQ)} & Ordinal, \newline 1 (Never felt) \dots\newline 10 (Str. felt)\\[2.0em]

\multirow{3}{4.8cm}{Please rate your agreement with the following statements w.r.t.\ both the VR sessions witnessed (Presence)} & Likert, \newline -3 (Str. Disagree) \dots\newline +3 (Str. Agree)\\[2.0em]

\multirow{3}{4.8cm}{Briefly tell us about your virtual experiences (suggestions/comments)} & \multirow{2}{3.2cm}{Free Response}\newline \\[2.0em]

\multirow{3}{4.8cm}{Extended comments/suggestions} & Free Response\\[2.0em] \bottomrule
\end{tabular}
\caption{Pre-experience and post-experience survey}
\label{tab:questions}
\end{table}

\begin{figure}[]
 \centering 
 \includegraphics[width=\columnwidth, trim={0 .27in 0 1.08in}, clip]{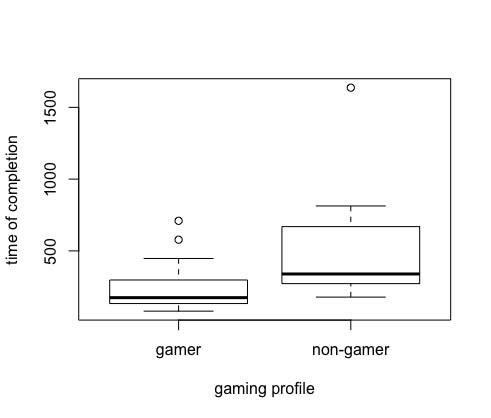}
 \caption{Time of completion is significantly less for participants who self-reported their gaming profile (gprofile) as gamers}
 \label{fig:rplot1}
\end{figure}

\begin{figure}[]
 \centering 
 \includegraphics[width=\columnwidth, trim={0 .27in 0 1.08in}, clip]{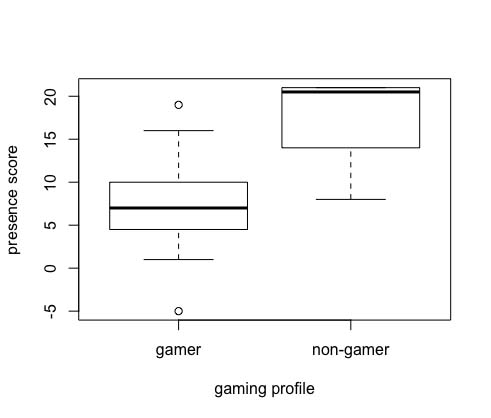}
 \caption{Presence score is higher for participants who self-reported their gaming profile (gprofile) as non-gamers}
 \label{fig:rplot3}
\end{figure}

\subsection{Procedure and Tasks} 
At the start of the experiment, participants were given elaborate verbal instructions explaining the experimental procedure. Both the gamer and the non-gamer group were then interviewed in order to profile their background and account for prior exposure to video gaming. The participants were given an overview of the technology involved and were explained how to wear a HMD and control their in-game avatar using the controller. They were also explained their responsibility towards the study (finding the targets inside the maze in a specific order, and if they were feeling nauseated, then they should immediately stop their VR experience). The participants were explicitly told that there were no time limits to their sessions and were advised to take their time to solve the maze. The participants would then initiate their two session maze experience with either immersive or non-immersive user interface control schematics. During the course of solving the maze, the participants had the option of verbally communicating with the study investigator in case they needed any further information.

\begin{figure}[t]
 \centering 
 \subfloat[3D mannequin model]{\includegraphics[width=40mm,scale=0.9]{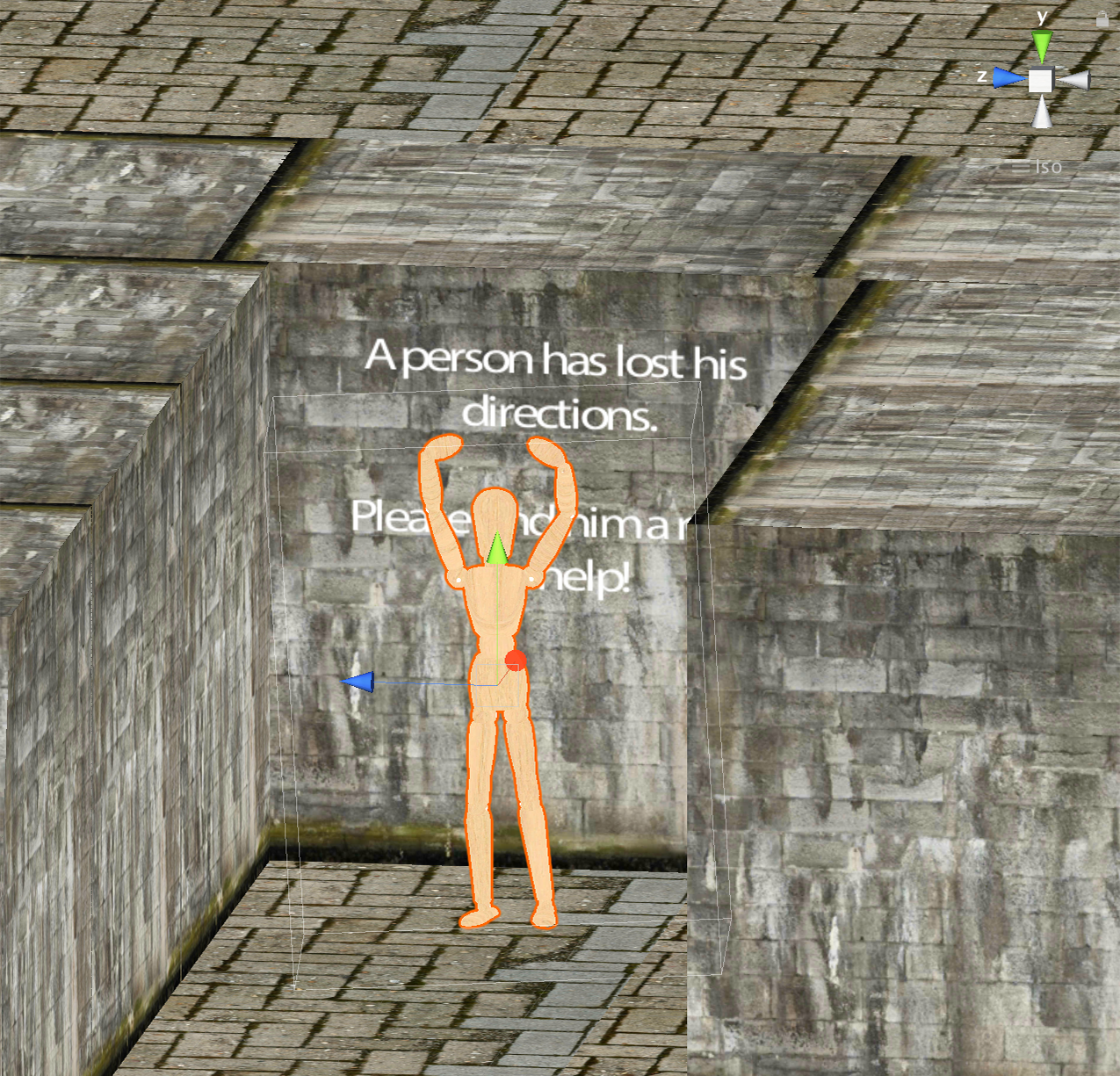}\label{fig:Avatar}}
 \hfill
 \subfloat[3D Minimap model]{\includegraphics[width=40mm,scale=0.9]{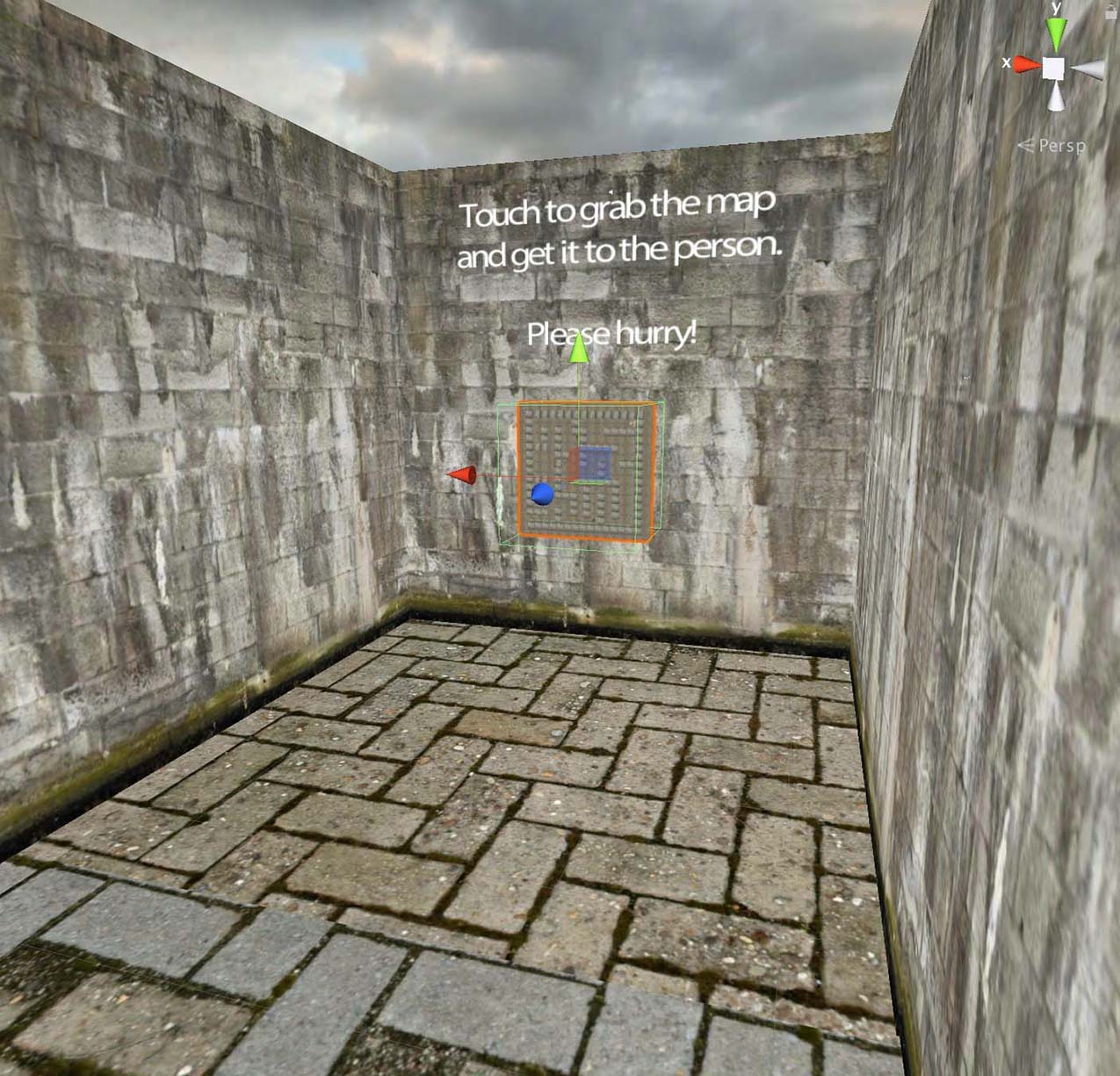}\label{fig:Minimap}}
 \caption{Targets inside the maze}
\end{figure}

The maze task model consisted of three tasks, all of which involve wayfinding. These tasks were as follows:
\begin{enumerate}
 \item Find your way to target 1, a mini map (\autoref{fig:Minimap})
 \item Find your way to target 2, the animated human avatar (\autoref{fig:Avatar})
 \item Find your way back to the entrance of the maze
\end{enumerate}

The difference between immersive and non-immersive user interface lies in the way the participants controlled their gaze. Immersive user interface control meant that the participants used their natural head orientation to align their view inside the maze environment (VE). Non-immersive user interface control meant the participants had to control their view inside the maze using the joystick analog control on the gaming controller (gamepad) (\autoref{fig:BothPerspective}). The participants were required to be seated while going through the study because this configuration helped minimize simulator sickness. After completing their first session, the participants were given a break of 5 minutes before starting their next session to help normalize any fatigue conditions. In their second session, they were given the same maze once again but with a different user interface control schematic from before. Upon completion of their second session, the post-experience survey was administered, which included a presence questionnaire \cite{usoh2000using} along with personal feedback. Simulator sickness questionnaires \cite{SSQ} were provided contextually in-between the sessions. At this point, all participant activity have been logged, collated and uploaded for further analysis.

\begin{figure}[th]
	\begin{center}
	   \includegraphics[width=\columnwidth]{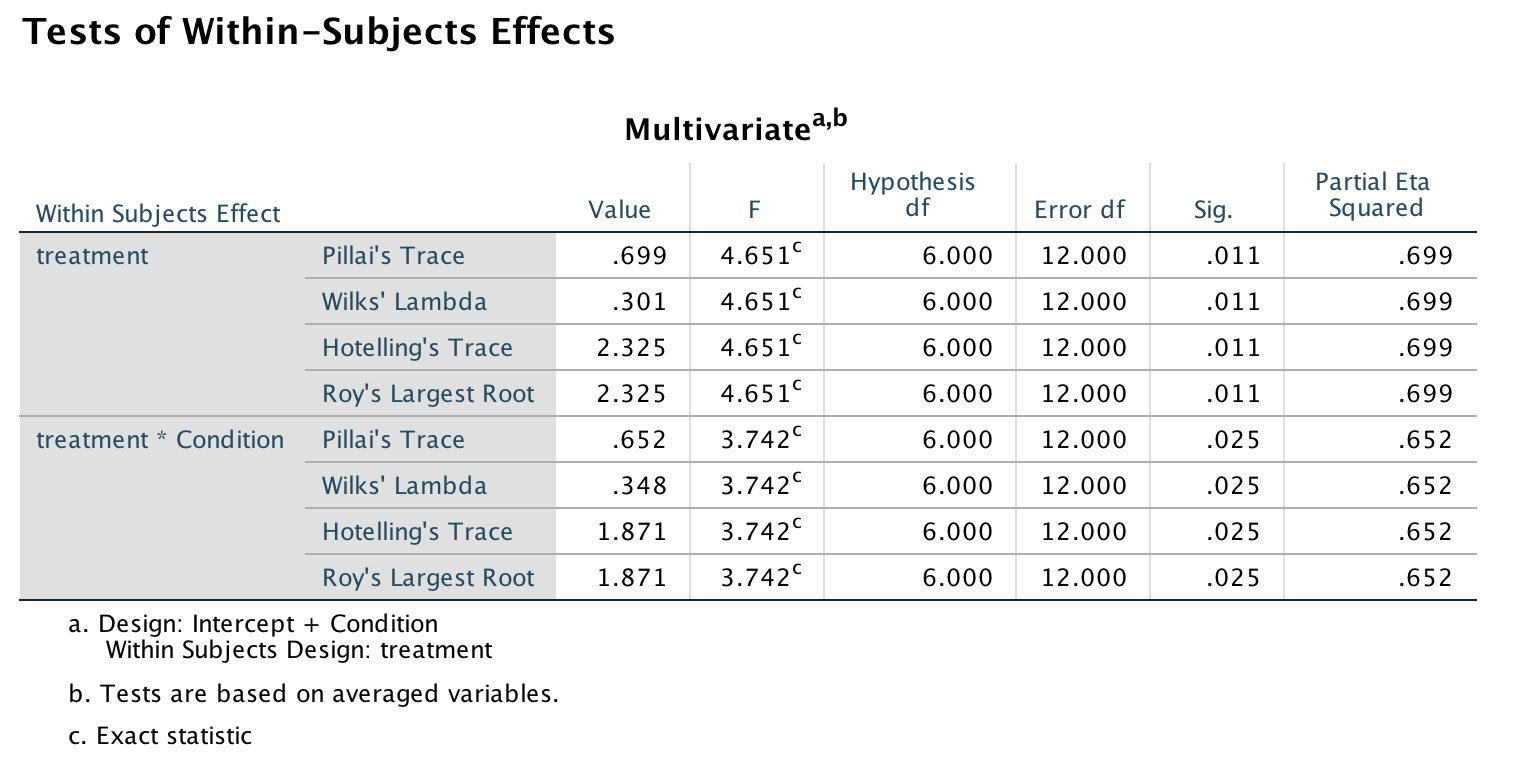}
	\end{center}
	\caption[Repeated measures analysis; test of within-subjects effects]
		{Repeated measures analysis for current user study; test of within-subjects effects.}
	\label{fig:ucavestudyfourrepeatedmeasureseffects}
\end{figure}

\begin{figure}[th]
	\begin{center}
	   \includegraphics[width=\columnwidth]{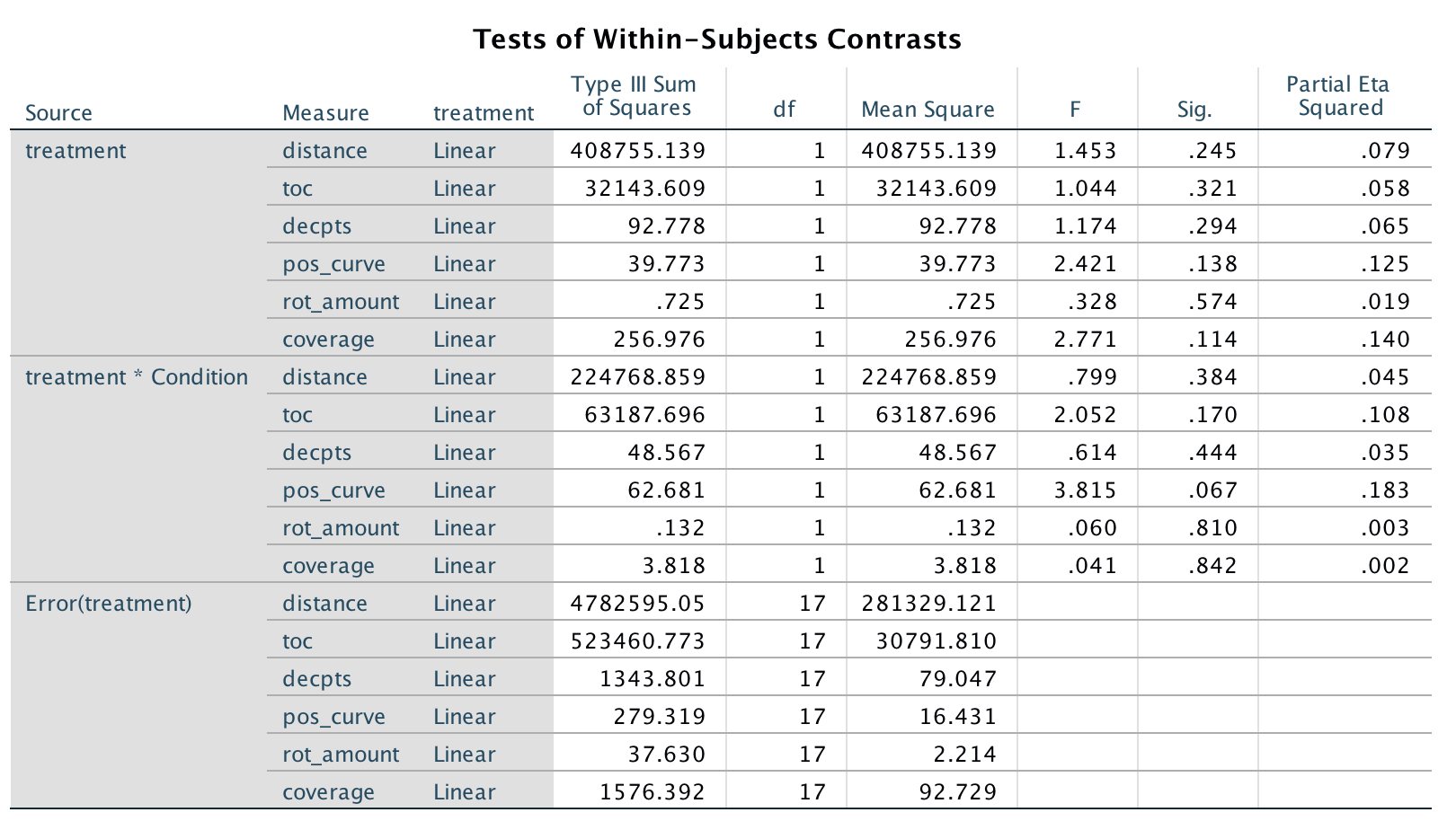}
	\end{center}
	\caption[Repeated measures analysis; test of within-subjects contrast]
		{Repeated measures analysis for current user study; test of within-subjects contrast.}
	\label{fig:ucavestudyfourrepeatedmeasurescontrast}
\end{figure}

\begin{figure}[th]
	\begin{center}
	   \includegraphics[width=\columnwidth]{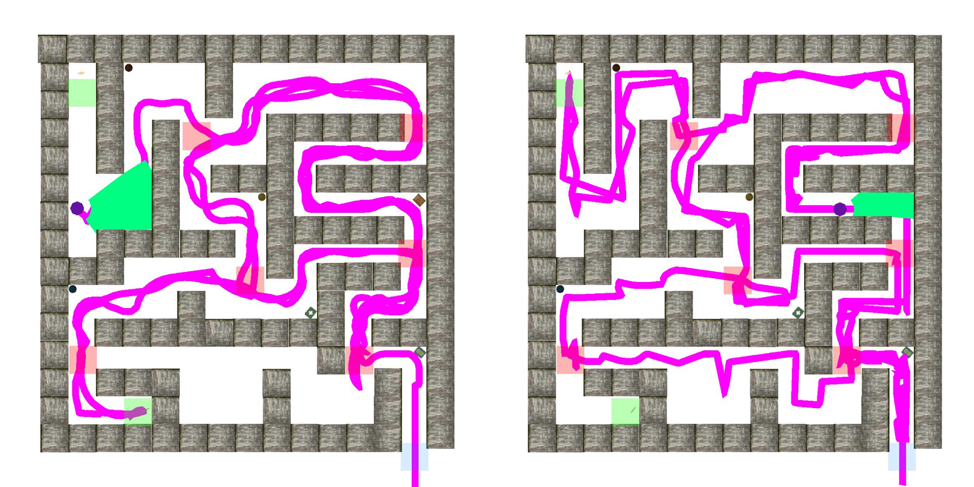}
	\end{center}
	\caption[Comparison of user trajectory while solving the maze]
		{Visual comparison of user trajectories of a randomly selected gamer (left) and a non-gamer (right) while solving the maze in the non-immersive interface.}
	\label{fig:ucavestudyfourtrajectorycomparison}
\end{figure}

\section{Results}\label{sec:results}
Only 19 participants out of 40 completed both maze sessions. We report that 2 participants out of 40 were not appropriately logged due to device I/O failure, resulting in complete data loss of 2 data points from the analysis. We investigate our dependent variable (time of completion) against the set of independent variables such as the participant's gaming profile type (gamer or non-gamer), the type of deployment (immersive and non-immersive), presence score, simulator sickness score, age, and gender.

\begin{figure}[th]
 \centering 
 \includegraphics[width=\columnwidth, trim={0 .32in 0 1.08in}, clip]{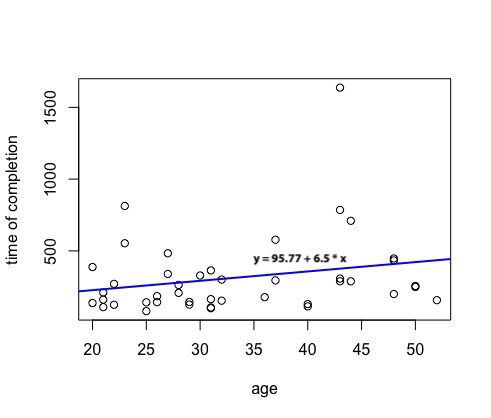}
 \caption [Maze Study - Scatterplot]
 {Maze Study - Graph showing scatterplot of time of completion of the maze as the y-axis and age of the participants as the x-axis.}
 \label{fig:rplot5}
\end{figure}

\begin{figure}[th]
 \centering 
 \includegraphics[width=\columnwidth, trim={0 .32in 0 1.08in}, clip]{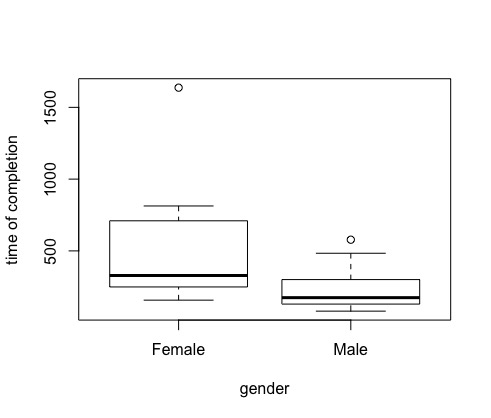}
 \caption [Maze Study - Time of completion of maze plotted against gender of the user]
 {Maze Study - Boxplot showing time of completion of the maze by group (gender).}
 \label{fig:rplot6}
\end{figure}

We report a significant difference between groups of means of gaming profile type and time of completion (F-value = 12.885, P = 0.000914). The box plot of time of completion with respect to gaming profile [see \autoref{fig:rplot1}] illustrates the difference between gamer and non-gamer groups. We observe, a majority of gamer participants took less time in solving the maze on average, whereas a majority of non-gamer participants took more time. Furthermore, figure \ref{fig:rplot5} shows a positive correlation (scatterplot) between the time of completion of the maze (in seconds) and the age of the participants. We also report that female participants in this study spent higher time completing the maze on average than the male participants in this study [see Figure \ref{fig:rplot6}]. This suggests that there is a trend of age and gender together impacting immersive VR user performance. 

\begin{figure}[]
 \centering 
 \includegraphics[width=\columnwidth, trim={0 .32in 0 1.08in}, clip]{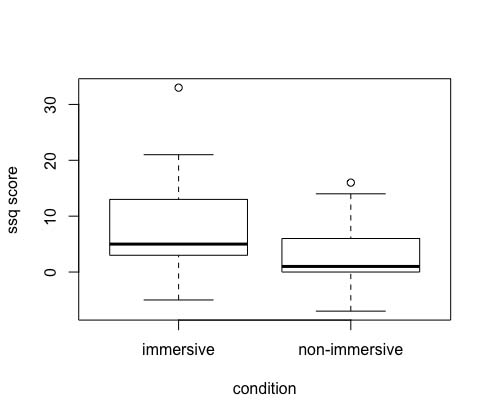}
 \caption{SSQ score is higher in immersive user interface control ~~~~~}
 \label{fig:rplot4}
\end{figure}

Additionally, we report a significant difference between groups of means of gaming profile type and self-reported presence score of the successful participants (F-value = 9.565, P = 0.00699). The box plot of presence score with respect to gaming profile [see \autoref{fig:rplot3}] illustrates the difference between gamer and non-gamer groups. Interestingly, we observed another significant difference between groups of means of condition of deployment and the self reported SSQ scores [see \autoref{fig:rplot4}]. It is indicative of the fact that immersive user interface control made participants nauseated to a higher degree than non-immersive user interface control. All of these findings together are indicative of a trend, but we need inferential statistics to make an argument conclusively. We then proceed to explore our recorded user trajectory dataset.

Repeated measures analysis of the user trajectory dataset revealed that technical factors, such as immersive and non-immersive user interfaces, do significantly impact immersive VR user performance [see Figure \ref{fig:ucavestudyfourrepeatedmeasurescontrast} and Figure \ref{fig:ucavestudyfourrepeatedmeasureseffects}]. This shows a significant main effect of treatment (difference between immersive and non-immersive user interface) and an interactive effect of treatment and condition (i.e. order of deployment; immersive or non-immersive first, immersive or non-immersive second) combined. We find that condition is not a significant factor between subjects. Looking closely at individual measures, we don't find a difference on any particular measure [see Figure \ref{fig:ucavestudyfourrepeatedmeasurescontrast}]. This suggests small but consistent differences that arise as a result of treatment (immersive versus non-immersive user interface).

\begin{figure}[t]
 \centering 
 \includegraphics[width=\columnwidth, trim={0 .32in 0 1.08in}, clip]{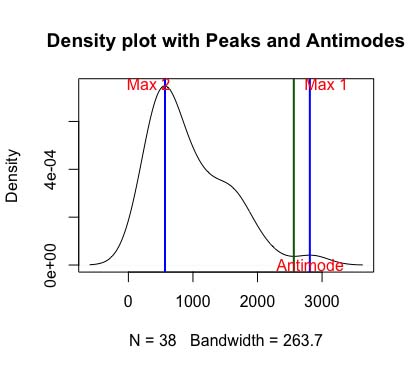}
 \caption{Bi-modality test}
 \label{fig:rplot8}
\end{figure}

From our DTW analysis, we found that the alignment distance of the trajectories (vector based feature) between the gamer group and non-gamer group is closer (tends to zero) in the case of immersive user interface than non-immersive user interface. This fact is further backed by a positive bi-modality test (see \autoref{fig:rplot8}) between the two groups. This suggests that immersive user interface normalizes the maze solving experience between the gamer and non-gamer groups.

\section{Discussion}\label{sec:discussion}
We \textit{reject} our null hypothesis and accept the alternate hypothesis. We found that prior exposure to video-gaming, age, and gender has a measurable impact on immersive VR user performance. While conceptualizing our experimental design, there were concerns about choosing the appropriate measurement parameters as performance metric for VR tasks, we worried that the immersive and non-immersive user interface control paradigms would be confusing to users, and that visual cues to understand decision making in mazes during analysis would be complicated and confounding. During the course of the study, we observed that users, who claimed to be self-perceived video-gamers, aren't confused by the immersive, non-immersive interface control paradigm. The video-gaming group seemed to treat both immersive and non-immersive user interface as the same, unlike the non-video-gaming group. The ability to replay user trajectory in the Unity game engine's editor environment proved to be very useful. These visual replays helped us determine user trajectory features such as, curvature, and head-rotation amount [see Figure \ref{fig:featurestrajectory}] to analyze. 

\par \textbf{Additional findings}: The interface worked well except twice when it failed to record the session due to a device malfunction. Due to the minimalist nature of our system setup, we were able to reduce points of failure in our system so that our participants could focus better on the study objectives. All user data logging was done in the background and were stored locally on-board the HMD device for faster I/O. The maze provided ample space-time interactivity for the participants which in turn provided us a rich data set including high resolution user trajectory data which can be parsed through in real-time [see Figure \ref{fig:sampleTrajectory}]. The task model in our study also proved to be fairly difficult for most of the participants, which helped offset the balance between chance-based decisions and skill-based decisions inside the maze. The average time that the users spent in immersive interface was 268 seconds while the average time spent by the same users in non-immersive interface was 364 seconds. As suspected, there were differences in performance between the gamer and the non-gamer groups [see Figure \ref{fig:ucavestudyfourtrajectorycomparison}]. In the context of spatial problem-solving, wayfinding, the combined finding of simulator sickness and presence score could be indicative of the fact that the two groups approached the study differently. From a pedagogical point of view, if immersive VR applications were to be deployed while making a meaningful impact to its user's skill-building exercise, we would have to account for the factor that different groups of users conceive and treat immersive environments differently based on their prior experience. Standardized immersive experience, although logistically good, is not effective in targeting individual specific needs. 

\section{Conclusions}\label{sec:concluded}
Immersive VR applications have been traditionally deployed in a controlled laboratory setting ever since the advent of HMDs \cite{Sutherland1968}. With the proliferation of highly accessible mobile communication devices such as the smartphone, high pixel density display technology became available as an alternative to dedicated and tethered HMDs. The mobility of these smart devices inspired us to conceive a mobile VR platform that is untethered and can be deployed universally. Our work represents another step toward ubiquitous deployment of immersive virtual experiences which will potentially advance general VR usability, incite more research, and further the cause for practical VR applications.

With the current study (a study of technical factor), we showed that prior exposure to video-gaming, age, and gender is found to have a measurable impact on user performance when it comes to spatial maze navigation in immersive VR. The current study design involved a relatively complicated maze design than Study III. The context of the maze experience was similar to Study III, in which the subject is required to find and rescue another human (avatar). The task included finding a clue (minimap) before finding the human avatar and then tracing back the path to the entrance of the maze. The hardware setup for the current study was kept similar to Study III for maintaining consistency. The current study design also looked at each participant partaking the maze experience twice with the distinction that one of the session would require the user to use only the game-controller to navigate the maze while the other would involve the same user solving the same maze combining natural head gaze rotation and the game-controller to navigate the maze. We found that the gaming profile coupled with age and gender is proving to be a strong indicator of success when it comes to navigating immersive mazes in VR. After our preliminary finding, we focused on the important issue of exploring the core impact on user task performance in an immersive VR setup deployed under an immersive setting. We conducted extended analysis on recorded user trajectory data. To that end, we defined a set of mathematically derived features generalized for each trajectory such as distance traveled, decision points reached inside the maze, positional curvature, head rotation amount, and coverage of the maze. Upon closer inspection of the variation in these trajectory features through repeated measure analysis, we found a small but consistent difference that arises as a result of the treatment of our final study design, that is immersive versus non-immersive user interface. We find that immersive user interface helps users explore the VE better with natural head gaze control than the non-immersive user interface with analog gaze control. These findings are useful to VR designers in making appropriate trade-offs in VR level design for specific VR applications.

To summarize, in this work, we have showed that prior exposure to video-gaming, age, and gender is found to have a measurable impact on user performance when it comes to spatial maze navigation in immersive VR. We conducted extended analysis on recorded user trajectory data. To that end, we defined a set of mathematically derived features generalized for each trajectory such as distance traveled, decision points reached inside the maze, positional curvature, head rotation amount, and coverage of the maze. Upon closer inspection of the variation in these trajectory features through repeated measure analysis, we found a small but consistent difference that arises as a result of the treatment of our final study design, that is immersive versus non-immersive user interface. We find that immersive user interface helps users explore the VE more with natural head gaze control than the non-immersive user interface with analog gaze control.


\bibliographystyle{acm}
\bibliography{references}


\end{document}